\documentstyle[12pt]{article} 
\begin{document}

\def\be{\begin{equation}}
\def\ee{\end{equation}}

\begin{flushright}
HRI-RECAPP-2009-005
\end{flushright}

\begin{center}
{\Large\bf A Randall-Sundrum scenario with bulk dilaton and torsion }\\[20mm]
Biswarup Mukhopadhyaya\footnote{E-mail: biswarup@mri.ernet.in}\\
{\em Regional Centre for Accelerator-based Particle Physics\\
Harish-Chandra Research Institute\\
Chhatnag Road, Jhusi, Allahabad - 211 019, India} 

Somasri Sen \footnote{E-mail: somasri.ctp@jmi.ac.in}\\
{\em Centre for Theoretical Physics\\
Jamia Millia Islamia\\
New Delhi 110 025, India}

Soumitra SenGupta \footnote{E-mail: tpssg@iacs.res.in} \\
{\em Department of Theoretical Physics\\
 Indian Association for the 
Cultivation of Science\\
Kolkata - 700 032, India}\\
\vspace{0.1cm}
\end{center}

\begin{abstract}
We consider a string-inspired torsion-dilaton-gravity action in a
Randall-Sundrum brane world scenario and show that, in an effective 
four-dimensional theory on the visible brane, the rank-2 antisymmetric
Kalb-Ramond field (source of torsion) is exponentially suppressed. The
result is similar to our earlier result in \cite{ourprl}, where no
dilaton was present in the bulk. This offers an explanation of the
apparent invisibility of torsion in our space-time.  However, in this
case the trilinear couplings $\sim$ Tev$^{-1}$ between dilaton and
torsion may lead to new signals in Tev scale experiments, bearing the
stamp of extra warped dimensions.

\end{abstract}

\vskip 1 true cm

\newpage
\section{Introduction}
Braneworld models with extra spatial dimensions have gained
considerable interest over the last decade, largely due to their
role in solving the naturalness/hierarchy problem in the context of
the standard model of elementary particles. Extra spacelike dimensions
with warped geometry elegantly bridges the gap of sixteen orders of
magnitude between electroweak and the Planck scale. The key feature of
the Randall-Sundrum type of theory\cite{rs} is the `$-$' exponential
warp factor attached to the Minkowski part of the metric, leading
to a suppression factor for all masses and gravitational couplings of
known fields residing on the visible brane located at one of the
orbifold fixed points of the extra compact spacelike dimension.

The essential input of the RS\cite{rs} type of theory is that 
gravity propagates in a 5-dimensional anti-de Sitter bulk spacetime,
while the standard model (SM) fields are assumed to lie on a
3-brane. String theory provides a natural explanation for such
description as the SM fields arise as excitations of an open string 
whose ends are fixed on the brane. The graviton, 
on the other hand, is a closed string
excitation and hence can propagate in the bulk
space-time\cite{string}.  Consequently, when one takes its projections
on the visible brane, the massless graviton mode has a coupling $\sim
1/M_P$ with all matter, while the massive modes have enhanced coupling
through the warp factor. The massless graviton mode therefore accounts
for the presence of gravity in our universe, while the massive modes
raise hopes for possible new signals of extra warped dimension in
accelerator experiments \cite{rssig}. However, in a string-based model,
one should also consider other closed string massless excitations
which appear as various higher rank tensor fields in the low energy
effective theory\cite{string}. They are expected to propagate in the
bulk just as gravity. Two such massless closed string excitations are
scalar dilaton and second rank antisymmetric Kalb Ramond (KR) field
\cite{burges}.The question we ask here is: can these fields have any
observable effects? If not, why are the effects of their massless
modes less perceptible on our brane than the force of gravitation? 
Implications of other type of bulk fields such as scalars\cite{gw},fermions
\cite{davod1} and gauge fields\cite{davod2} have also been considered in
this scenario. While some  such scenarios are testable in
accelerator experiments \cite{bulkac} or in the neutrino sector
\cite{gross}, by and large they do not cause any contradiction with
our observations so far.
 
The antisymmetric rank-2 tensor bulk field namely the
Kalb-Ramond\cite{kr} field has similar coupling to matter as
gravity. Using a generalised form of the Einstein-Cartan action, it
has been shown earlier that the third rank field strength tensor of KR
field can be identified with torsion \cite{ssgpm}. Experimental limits
impose severe constrains on such fields so that it is
vanishingly small\cite{torlim}. This apparent invisibility of torsion
has been clarified in one of our earlier work \cite{ourprl}. There it
has been shown that the zero mode of the antisymmetric tensor field
gets an additional exponential suppression compared to the graviton on
the visible brane. This could well be an explanation of why we see the
effect of curvature but not of torsion in the evolution of the
universe. But as mentioned earlier, string theory allow dilaton as
well to exist in the bulk and hence it is essential to consider the
dilaton-torsion coupling in the bulk, instead of taking a free torsion
field. In this light, the earlier conclusion regarding the suppression
of the torsion field on our brane must be re-examined when it is
coupled to the dilaton in the bulk. We address this question in the
current study in context of RS braneworld scenario. We also explore
other possible implications of bulk dilaton field on the Physics on
our brane. It may be mentioned that the issue of braneworld stability
in a similar dilaton-torsion coupled scenario has been investigated by
others\cite{sdssg}.

We organise our work as follows. In the next section, we review the RS
theory and then illustrate our model in the background of RS
spacetime, where torsion coexist with gravity and dilaton in the bulk
and has a non-minimal coupling with dilaton as one obtains in string
inspired models. Section 3 presents the solution of the bulk dilaton
on the visible brane. In section 4, we describe in detail how the
different KK modes of the KR field are coupled to the dilaton fields
on the brane, and how these couplings affect our previous explanation
of the invisibility of torsion on our SM brane. From this, we conclude
that even in presence of dilaton coupling the zero mode of the KR
field is suppressed with an additional exponential
suppression. Thus this work reinforces our claim that RS scenario can
provide an explanation of why the effect of torsion is much weaker
than that of curvature on the brane. However, because of the presence
of the dilaton, we obtain trilinear couplings between dilaton-KR field
which may have their imprints in Tev-scale Physics.

\section{Dilaton-Torsion in Randall Sundrum framework}

In the Randall-Sundrum model, we have a 5-dimensional
anti de-sitter spacetime with the extra spatial dimension orbifolded
as $S_1 /Z_2$. Two 4 dimensional branes known as the visible (TeV)
brane and hidden (Planck) brane are placed at the two orbifold fixed
points $y = 0$ and $\pi$ . The 5-dimensional spacetime is characterised
by the metric
\be
ds^2= e^{-2\sigma}\eta_{\mu\nu}dx^{\mu}dx^{\nu}+r_c^2 dy^2
\ee                                                              
where $\eta_{\mu\nu}$ is the usual 4D Minkowski metric with the sign
convention $(-,+,+,+)$, and $\sigma = kr_c|y|$. $r_c$ is the
compactification radius for the fifth dimension, and $k$ is on the
order of the higher dimensional Planck mass $M$ . (Greek indices are
used for brane coordinates and Latin indices are used for the full 5D
coordinates). The standard model fields reside at $y = \pi$ while
gravity peaks at $y = 0$. The dimensionful parameters defined above are
related to the 4-dimensional Planck scale $M_P$ through the relation
\be
M_P^2=\frac{M^3}{k}[1-e^{-2kr_c\pi}]
\ee
Clearly, $M_P$ , $M$ and $k$ are all of the same order of
magnitude. As a consequence of warped geometry, all mass scales get
exponentially lowered from the Planck scale to the TeV scale because
of the warp factor $e^{−kr_c\pi}$. Thus the hierarchy between the
Planck and TeV scales can be explained without the need of any
fine-tuning.

In the scenario adopted by us the source of torsion is taken to be the
rank-2 anti symmetric KR field $B_{MN}$\cite{ssgpm}. It has been shown
in \cite{ssgpm}, that in the low energy effective string action the
torsion field $T_{MNL}$, which is an auxiliary field, could be
identified with the rank-3 anti symmetric field $H_{MNL}$ by using the
equation of motion $T_{MNL}=H_{MNL}$. Torsion can be identified with
$H_{MNL}$ which is the field strength tensor of the KR field $B_{MN}$.

In the context of string theory, apart from gravity, scalar dilaton
and the second rank anti-symmetric tensor KR field both co-exist in
the bulk. The five dimensional action for the gravity-dilaton-torsion
sector is given by
\be
S_{tot}=\int d^5x \sqrt{-G}[2M^3R
-e^{\phi/M^{3/2}}2H_{MNL}H^{MNL}+\frac{1}{2}\partial_M\phi\partial^M\phi-m^2
\phi^2]
\ee
 where 
\begin{eqnarray}
S_{grav} &=&\int d^5x \sqrt{-G} 2M^3R\\
S_{KR} &=& \int d^5x \sqrt{-G}[-e^{\phi/M^{3/2}}2H_{MNL}H^{MNL}]\\
S_{dil} &=& \int d^5x \sqrt{-G}[\frac{1}{2}\partial_M\phi\partial^M\phi-m^2 \phi^2]
\end{eqnarray}

From now onwards, we focus on the the part of the action consisting of
$S_{KR}$ and $S_{dil}$ only. Our aim is to find out the influence of
the dilaton field on the KR Lagrangian, and to see how it affects the
existence of different torsion modes on the visible brane. Moreover, we
take into account the massless dilaton field only, as the massive
field on the bulk will anyway decouple. The relevant part of the
action is,
\be
S=\int d^5x \sqrt{-G}[-e^{\phi/M^{3/2}}2H_{MNL}H^{MNL}+\frac{1}{2}\partial_M\phi\partial^M\phi]
\ee
Using the explicit form of the background metric in the Randall-Sundrum scenario
and also the KR gauge fixing condition $B_{4\mu}=0$, the above action can be written as
\begin{eqnarray}
S=\int d^4 x dy
r_c[&-&2e^{\phi/M^{3/2}}\{e^{2\sigma}\eta^{\mu\alpha}\eta^{\nu\beta}\eta^{\lambda\gamma}\partial_{[\mu}B_{\nu\lambda]}\partial_{[\alpha}B_{\beta\gamma]}-\frac{3}{r_c^2}\eta^{\mu\alpha}\eta^{\nu\beta}B_{\mu\nu}\partial^2_y
B_{\alpha\beta}\}\nonumber\\
&+&\frac{1}{2}\{e^{-2\sigma}\eta^{\mu\nu}\partial_{\mu}\phi\partial_{\nu}\phi-\frac{1}{r_c^2}\phi\partial_y[e^{-4\sigma}\partial_y\phi]\}]
\end{eqnarray}

Next, consider Kaluza-Klein decomposition of the following form
for both the KR and the dilatonic field
\begin{eqnarray}
B_{\mu\nu}=\sum_{n=0}^{\infty}B^n_{\mu\nu}(x)\frac{\chi^n(y)}{\sqrt{r_c}}\\
\phi=\sum_{n=0}^{\infty}\phi^n(x)\frac{\psi^n(y)}{\sqrt{r_c}}
\end{eqnarray}
In terms of the projections $(B^n_{\mu\nu}(x),\phi^n(x))$ on the
visible brane our effective action looks like
\begin{eqnarray}
S=\int d^4 x [\sum_n\sum_m \int dy e^{2\sigma}\chi^n\chi^m \{
-2e^{\frac{1}{M^{3/2}}\sum_i\frac{\phi^i\psi^i}{\sqrt{r_c}}}
\eta^{\mu\alpha} \eta^{\nu\beta} \eta^{\lambda\gamma}
\partial_{[\mu}B^n_{\nu\lambda]} \partial_{[\alpha}
B^m_{\beta\gamma]}\nonumber\\-3[\frac{1}{r_c^2}\frac{e^{-2\sigma}}{\chi^m}
\partial^2_y\chi^m]\eta^{\mu\alpha}\eta^{\nu\beta}B^n_{\mu\nu}B^m_{\alpha\beta}\}]\nonumber\\
+\frac{1}{2}[\sum_n\sum_m \int dy
e^{-2\sigma}\psi^n\psi^m\{\eta^{\mu\nu}\partial_{\mu}\phi^n\partial_{\nu}\phi^m-[\frac{1}{r_c^2}\frac{e^{2\sigma}}{\psi^m}\partial_y(e^{-4\sigma}\partial_y\psi^m)]\phi^n\phi^m\}]
\end{eqnarray}
This essentially represents the four dimensional action where the term
in the first square bracket is the lagrangian corresponding to the KR
field coupled to dilaton and the second square bracketed term  is the lagrangian 
for a dilaton field. It is interesting to note that the second term
which is basically $S_{dil}$ has no coupling with the KR field and
hence corresponds to a free dilaton action. We first proceed with this
part to find out how the massless bulk dilaton field that appears on
the standard model brane.

\section{Bulk Dilaton Field} 
It is straightforward to check from above that the dilaton action
assumes a standard canonical form on the standard model brane i.e.,
\be
S_{dil}=\int d^4 x \sum_{n=0}^{\infty} \frac{1}{2} (\eta^{\mu\nu}\partial_\mu\phi^n\partial_\nu\phi^n+{m_n^d}^2 {\phi^n}^2)
\ee
provided $m_n^d$, the mass of the nth mode of the dilaton field, is defined by  
\be
-\frac{1}{r_c^2}\frac{e^{2\sigma}}{\psi^m}\partial_y(e^{-4\sigma}\partial_y\psi^m)]={m_n^d}^2 
\ee
and $\psi^m(y)$ satisfy the orthonormality relation 
\be
\int^{\pi}_{-\pi} dy e^{-2\sigma}\psi^n(y)\psi^m(y)=\delta_{nm}.
\ee
In terms of $z_n=\frac{m_n^d}{k}e^\sigma$ and $f^n=\psi^n e^{-2\sigma}$ equation (13)
can be recast in the form
\be
z_n^2\frac{d^2f^n}{dz_n^2}+z_n\frac{df^n}{dz_n}+[z_n^2-4]f^n=0
\ee
The above equation admits a solution of Bessel function of order
2. Consequently, $\psi^n$ has a solution of the form
\be
\psi^n=\frac{e^{2\sigma}}{N_n}[J_2(z_n)+\alpha_n Y_2(z_n)]
\ee 
where $J_2(z_n)$ and $Y_2(z_n)$ are Bessel and Neumann functions of
order 2, $\alpha_n$ and $N_n$ are two constants to be determined from
the boundary conditions at the orbifold fixed points.  The continuity
condition for the derivative of $\psi^n$ at $y=0$, dictated by the
self-adjointness of the left hand side of equation(13), yields
\be
\alpha_n=-\frac{J_1(\frac{m_n^d}{k})}{Y_1(\frac{m_n^d}{k})}
\ee
Using the fact that $e^{kr_c\pi}>>1$ and the mass values $m_n^d$ on
the brane is expected to be on the order of TeV scale $(<<k)$,
\be
\alpha_n=-\frac{\pi}{4}(\frac{m_n^d}{k})^2<<1
\ee
The boundary condition at $y=\pi$ gives
\be
J_1 (x_n ) = 0
\ee
where $x_n=z_n(\pi) = \frac{m_n^d}{k}e^{kr_c\pi}$.\\ The roots of the
above equation determine the masses of the higher excitation modes. As
$x_n$ is of order unity, the massive modes lie in the TeV scale. The
point to note here is that the masses of the higher excitation modes
of the dilaton field are same as the massive graviton modes. 

The orthonormality condition determines the constant $N_n$:
\be
N_n=\frac{e^{kr_c\pi}}{\sqrt{kr_c}}J_2(x_n)
\ee
Thus the final solution for the massive modes turns out to be
\be
\psi^n(z_n)=\sqrt{kr_c}\frac{e^{2\sigma}}{e^{kr_c\pi}}\frac{J_2(z_n)}{J_2(x_n)}
\ee
We shall compare the lowest-lying massive modes of the dilaton field
with those of graviton as well as the bulk KR field in the next
section, when we determine the masses for the latter.

The solution for the massless dilaton mode can be obtained from
equation (13) for $m_n^d=0$ and is given by,
\be
\psi^0 =\frac{C_1}{4krc}e^{4\sigma}+C_2
\ee
With the help of the same boundary conditions that has been used for
the massive case, to determine the constants $(C_1,C_2)$, one finally
arrives at the solution for the massless mode as,
\be
\psi^0=\sqrt{kr_c}
\ee 
The complete solution for the dilaton field is then, 
\be
\phi(x,y)=\sqrt{k}\phi^0(x)+\sum_{n=1}^{\infty}\sqrt{k}\frac{e^{2\sigma}}{e^{kr_c\pi}}\frac{J_2(z_n)}{J_2(x_n)}\phi^n(x)
\ee
On our visible brane the field would appear as
\be
\phi(x,y=\pi)=\sqrt{k}\phi^0(x)+\sum_{n=1}^{\infty}\sqrt{k}e^{kr_c\pi}\phi^n(x)
\ee
At this point it is useful to mention that the solution for the bulk
massless dilaton is the same as that for the bulk graviton
field. Unlike the massless case, the massive modes in both the cases
appear with an exponential enhancement factor.

\section{Bulk KR field coupled to dilaton}

In this section we proceed with the term in the first square bracket in
equation(11) where the dilaton field is coupled to the bulk KR
field. Since we have already found out how the bulk dilaton field
would appear on the visible brane (eqn(12)) and what would be the
structure of the dilaton field (eqn(24)), we now use this
solution in our subsequent analysis.
\begin{eqnarray}
S_{KR}=\int d^4 x
[\sum_n\sum_m \int dy e^{2\sigma}\chi^n\chi^m \{ -2e^{\frac{\sqrt{k}}{M^{3/2}}[\phi^0+\sum_i\frac{e^{2\sigma}J_2(z_i)}{e^{kr_c\pi}J_2(x_i)}\phi^i(x)]} (\eta^{\mu\alpha} \eta^{\nu\beta} \eta^{\lambda\gamma} \partial_{[\mu}B^n_{\nu\lambda]} \partial_{[\alpha} B^m_{\beta\gamma]}\nonumber\\
-3[\frac{1}{r_c^2}\frac{e^{-2\sigma}}{\chi^m}
\partial^2_y\chi^m]\eta^{\mu\alpha}\eta^{\nu\beta}B^n_{\mu\nu}B^m_{\alpha\beta})\}]\\
=\int d^4 x
[\sum_n\sum_m \int dy e^{2\sigma}\chi^n\chi^m \{ -2e^{\frac{\sqrt{k}}{M^{3/2}}[\phi^0+\sum_i
b_i(y)\phi^i(x)]}(H^n_{\mu\nu\lambda}{H^m}^{\mu\nu\lambda}+3{m_m^t}^2 B^n_{\mu\nu}{B^m}^{\mu\nu})\}]
\end{eqnarray}
As discussed earlier, the third rank anti-symmetric field strength
$H^n_{\mu\nu\lambda}=\partial_{[\mu}B^n_{\nu\lambda]}$ may be
identified as torsion on our 4D brane,
$$b_i(y)=\frac{e^{2\sigma}J_2(z_i)}{e^{kr_c\pi}J_2(x_i)}$$
 and
$$-\frac{1}{r_c^2}\frac{e^{-2\sigma}}{\chi^m}\partial^2_y\chi^m={m_m^t}^2$$
defines the mass of the $m$th mode of the KR field. This equation is
much like equation (13). In terms of
$z_m^\prime=\frac{m_m^t}{k}e^\sigma$ the equation can be recast in the
form
\be
{z_m^\prime}^2\frac{d^2\chi^m}{d{z_m^\prime}^2}+z_m^\prime\frac{d\chi^m}{dz_m^\prime}+{z_m^\prime}^2\chi^m=0
\ee
This is a Bessel equation and solutions of this equation are Bessel
functions of order $0$.
\be
\chi^m=\frac{1}{N_m^\prime}[J_0(z_m^\prime)+\alpha_m^\prime Y_0(z_m^\prime)]
\ee
where $J_0(z_m^\prime)$ and $Y_0(z_m^\prime)$ are Bessel and Neumann
functions of order 0 and $\alpha_m^\prime$ and $N_m^\prime$ are two
constants in this case to be determined from the boundary
conditions. The continuity condition for the derivative of $\chi^m$ at
$y=0$ yields
\be
\alpha_m^\prime=-\frac{J_0^\prime(\frac{m_m^t}{k})}{Y_0^\prime(\frac{m_m^t}{k})}=-\frac{J_1(\frac{m_m^t}{k})}{Y_1(\frac{m_m^t}{k})}
\ee
Here
$J_0^\prime(\frac{m_m^t}{k})=\frac{d}{dz_m^\prime}J_0(z_m^\prime)|_{y=0}$. Like
the previous section we use the fact that $e^{kr_c\pi}>>1$ and the
mass values $m_m^t$ on the brane are expected to be in the TeV
range $(<<k)$,
\be
\alpha_m^\prime=-\frac{\pi}{4}(\frac{m_m^t}{k})^2<<1
\ee
The continuity condition for the derivative of $\chi^m$ at $y=\pi$
determines $m_m^t$ from the roots of the equation
\be
J_1 (x_m^\prime) = 0
\ee
where $x_m^\prime=z_m^\prime(\pi) = \frac{m_m^t}{k}e^{kr_c\pi}$. It is
interesting to see that the masses of the excited modes for both the
KR field and dilaton field are governed by the same equation. While
the masses of the higher modes of the dilaton field are the same as
those of the graviton modes, the massive KR fields are scaled by a
factor of $\sqrt3$ from the other two. We enlist the masses of a few 
low-lying modes of graviton, dilaton and KR field in table 1.
\begin{center}
$$
\begin{array}{|c|c c c c|}
\hline
n  & 1  & 2  & 3  & 4  \\
\hline
m_n^{grav}~(TeV) & 1.66 & 3.04 & 4.40 & 5.77 \\
\hline
m_n^{dil}~(TeV) & 1.66 & 3.04 & 4.40 & 5.77 \\
\hline
m_n^{KR}~(TeV)  & 2.87 & 5.26 & 7.62 & 9.99 \\
\hline

\end{array}
$$ 
\end{center} 

{\em Table 1: The masses of a few low-lying modes of the graviton, dilaton
   and Kalb-Ramond fields, for $kr_c=12$ and $k=10^{19}$GeV.}

\vspace{0.2cm}

So, as a result of these boundary conditions $\chi^m$ looks like
\be
\chi^m=\frac{1}{N_m^\prime}J_0(z_m^\prime)
\ee
where $N_m^\prime$ is yet to be determined. Now we go back to
equation(27) and make a series expansion of the exponential term
\begin{eqnarray}
S_{KR}&=&\int d^4 x\sum_n\sum_m [\int dy e^{2\sigma}\chi^n\chi^m
[1+\frac{\sqrt{k}}{M^{3/2}}\left(\phi^0+\sum_i b_i(y)\phi^i\right)\nonumber\\
&+&\frac{1}{2}\left(\frac{\sqrt{k}}{M^{3/2}}\right)^2
\left[{\phi^0}+(\sum_i b_i(y)\phi^i)\right]^2+........]
\{ -2(H^n_{\mu\nu\lambda}{H^m}^{\mu\nu\lambda}+3{m_m^t}^2 B^n_{\mu\nu}{B^m}^{\mu\nu})\}
\end{eqnarray}
The first term of the series in the integral represent solely the
lagrangian for the free KR field in 4D. The integral on the extra
dimension, associated with this term, defines the orthogonality
relation for the function $\chi^n$ as,
\be
\int dy e^{2\sigma}\chi^n\chi^m=\delta_{mn}
\ee
This is the same as found in ref\cite{ourprl}. The orthogonality
relation determines the normalisation constant $N_m^\prime$ as
\be 
N_n^\prime=\frac{e^{kr_c\pi}}{\sqrt{kr_c}}J_0(x_n^\prime)
\ee
Using this the final solutions for the massive modes for the KR field
turns out to be
\be
\chi^n(z_n^\prime)=\sqrt{kr_c}{e^{-kr_c\pi}}\frac{J_0(z_n^\prime)}{J_0(x_n^\prime)}
\ee
The solution for the massless mode of the KR field can be obtained by
solving the differential equation for $m_m^t =0$. This yields,
\be 
\chi^0=C_3|y|+C_4
\ee
The condition of self-adjointness implies the solution will be a
constant which can be determined from the normalisation condition as,
\be
\chi^0=\sqrt{kr_c}e^{-kr_c\pi}
\ee
This identically matches our solution in ref \cite{ourprl} and thereby
consolidates our claim that due to the presence of this large
exponential suppression factor the KR field is heavily suppressed on
the visible brane and thus makes the torsion imperceptible.

Let us now look back at equation (34). Different terms of the series
exhibit different couplings (trilinear and higher order) between the
KR field and dilaton field. The second term demonstrate a trilinear
coupling of the form $H^2\phi$. The coupling of the massless mode of
the dilaton field with the massless KR mode is suppressed by
$~\frac{1}{M_P}$ (k is taken to be on the order of Planck
scale). However, the trilinear coupling of the form $\phi^0 H^0 H^n$
vanishes due to the orthogonality relation. We have to figure
out the coupling scale of the massive modes of the dilaton field to
the KR field (both for its massive and massless modes). But the
subsequent terms of the series, representing quartic and higher order
coupling between the two fields, are even more heavily suppressed by the
factors of $~\frac{1}{M_P^2}$ or more.  Thus there is hardly any
phenomenological implication of such terms in the context of terrestrial 
experiments. 

Next, we calculate the coupling of
the massive dilaton fields with the KR fields on the brane. For this,
we use the solutions which we have obtained so far for both the
fields.
 
From equation (34), the strength of the trilinear coupling between the
$i$th massive mode of the dilaton field and the $n$th and $m$th mode
of the KR field is given by,
\be
\frac{\sqrt{k}}{M^{3/2}} \int_{-\pi}^{\pi} dy e^{2\sigma} \chi^n\chi^m  b_i(y)
\ee
Using the expression for $b_i(y)$ and the solution for massive mode of 
KR field $\chi^n$ as given in equation(37), the above integral becomes,
\be
\frac{kr_c}{M_P e^{3kr_c\pi}} \int_{-\pi}^{\pi} dy e^{4\sigma}\frac{J_0(z_n^\prime)J_0(z_m^\prime)J_2(z_i)}{J_0(x_n^\prime)J_0(x_m^\prime)J_2(x_i)}
\ee
while the coupling expression for the massive dilaton field with the massless KR field is
\be
\frac{kr_c}{M_P e^{3kr_c\pi}} \int_{-\pi}^{\pi} dy e^{4\sigma}\frac{J_2(z_i)}{J_2(x_i)}
\ee
We compute the integrals numerically taking $kr_c=12$ and considering
the mass of different modes of the fields from table 1. The couplings
for the all the massive fields turn out to be $0.99\times 10^{-3}$
GeV$^{-1}$ while the coupling for the massive dilaton with massless KR
is $7.51\times 10^{-3}$ GeV$^{-1}$. Thus the interaction strength is
suppressed by TeV$^{-1}$ and the lowest lying mass modes are also in
the TeV scale. This yields the rather interesting possibility: the KR zero mode,
which is shown to have extreme suppresed coupling with all matter fields,
can now be produced in TeV-scale hadron collider experiments through
the Drell-Yan process mediated by massive dilaton modes. Since the
interaction with gluons or quark-antiquark pairs are now suppressed
by only the TeV-scale, such production may have significant rates
for sufficient integrated luminosity, and may show up in final
states with missing transverse energy with some accompanying tagged
jets. In addition, massive KR modes, too, can in principle be 
pair-produced, leading to final states that may look akin to those
from graviton pairs, but may differ in angular distributions etc.
due to the characteristic Lorenz structure of the interactions 
involved. Thus dilaton mediation proves to be of interest, so far
as the experinmental signatures of the KR tower are concerned.

\section{Conclusions}
Extending our earlier work \cite{ourprl}, we have considered the full
bosonic sector of the low energy effective string action comprising of
bulk dilaton, KR and gravity embedded in a five dimensional RS
scenario, and have shown that the heavily suppressed nature of torsion
on our visible brane is unaffected by its non-trivial coupling with
the dilaton inside the bulk. This explanation of the torsion-free
nature of our space-time originates in the warped braneworld model
proposed by Randall and Sundrum.

We have also shown that such an warped extra dimensional model
indicates the presence of new interactions between the Kaluza Klein
modes of the dilaton and KR field which may be of significant
phenomenological importance.  The compactification of the fifth
dimension gives rise to a spectrum of Kaluza Klein modes for all the
fields including gravity. It is interesting to notice that the
condition that determines the masses of the different Kaluza Klein
modes of all the three fields are same, which eventually identifies
the same mass for different modes for graviton and dilaton while the
massive KR spectrum is scaled by a factor of $\sqrt{3}$ with respect
to the others[see table 1]. It is shown that the trilinear coupling
between dilaton and KR field are of the order of TeV$^{-1}$ and hence
can lead to detectable signals in TeV scale accelerator
experiments. The other higher order (quartic and so on) couplings are
again very highly suppressed $(~{\cal{O}}(\frac{1}{M_P^2}))$ and won't
have a visible impact.

\section{Acknowledgement}
We thank Ashoke Sen for helpful comments. SS acknowledges
the financial support provided by the University Grants' Commission
under Dr. D.S. Kothari fellowship scheme. The work of BM
was partially supported by funding available
from the Department of Atomic Energy, Government of India, for the
Regional Centre for Accelerator-based Particle Physics,
Harish-Chandra Research Institute.

\end{document}